\documentclass[5p]{elsarticle}
\usepackage[T1]{fontenc}
\usepackage{hyperref}
\usepackage{mathtools}
\usepackage{booktabs}
\usepackage{threeparttable}
\usepackage{siunitx}
\usepackage{amsmath}
\usepackage{amssymb}

\makeatletter
\def\ps@pprintTitle{%
 \let\@oddhead\@empty
 \let\@evenhead\@empty
 \def\@oddfoot{}%
 \let\@evenfoot\@oddfoot}
\makeatother
\begin{document}

\begin{frontmatter}

\title{Solenogam: A new detector array for $\gamma$-ray and conversion-electron spectroscopy of long-lived states in fusion-evaporation products}

%%Authors%%%%%%%%%%%%%%%%%%%%%%%%%%%%%%%%%%%%%%%%%%%%%%%%
\author[ANU]{M.S.M.~Gerathy\corref{cor1}} 
\ead{matthew.gerathy@anu.edu.au}

\author[ANU]{G.J.~Lane}
\author[ANU]{G.D.~Dracoulis\fnref{fn1}}
\author[ANU,FOR]{P.~Nieminen}
\author[ANU]{T.~Kib\'edi}
\author[ANU]{M.W.~Reed}
\author[ANU]{A.~Akber}
\author[ANU]{B.J.~Coombes}
\author[ANU]{M.~Dasgupta}
\author[ANU]{J.T.H.~Dowie}
\author[ANU]{T.J.~Gray}
\author[ANU]{D.J.~Hinde}
\author[ANU]{B.Q.~Lee}
\author[ANU]{A.J.~Mitchell}
\author[ANU]{T.~Palazzo}
\author[ANU]{A.E.~Stuchbery}
\author[ANU]{L.~Whichello}
\author[ANU]{A.M.~Wright}

\address[ANU]{Department of Nuclear Physics, Research School of Physics, The Australian National University, Canberra, ACT, 2601, Australia}
\address[FOR]{Fortum Power and Heat Oy, Generation Division, P.O. Box 100, 00048 FORTUM, Finland}

\cortext[cor1]{Corresponding author}
\fntext[fn1]{Deceased}

%%%%%%%%%%%%%%%%%%%%%%%%%%%%%%%%%%%%%%%%%%%%%%%%%%%%%%%%%%
\begin{abstract}
A new detector array, Solenogam, has been developed at the Australian National University Heavy Ion Accelerator Facility.  Coupled initially to the SOLITAIRE 6.5~T, gas-filled, solenoidal separator, and later to an 8~T solenoid, the system enables the study of long-lived nuclear states through $\gamma$-ray and conversion-electron spectroscopy in a low-background environment.  The detector system is described and results from the commissioning experiments are presented.
\end{abstract}


\end{frontmatter}
\sloppy

%---------------------------------------------------------------------------------------
\section{Introduction}

The study of isomeric states offers insight into nuclear structure, highlighting changes in the underlying behaviour of the nucleons.  Transition rates between nuclear states depend on changes in excitation energy, angular momentum, parity, and state character.  In particular, when there is a large change in the nuclear wavefunction, the associated transitions are inhibited, which can result in a relatively long lifetime.

\begin{figure*}[!htb]
\begin{center}
\includegraphics[width=\textwidth]{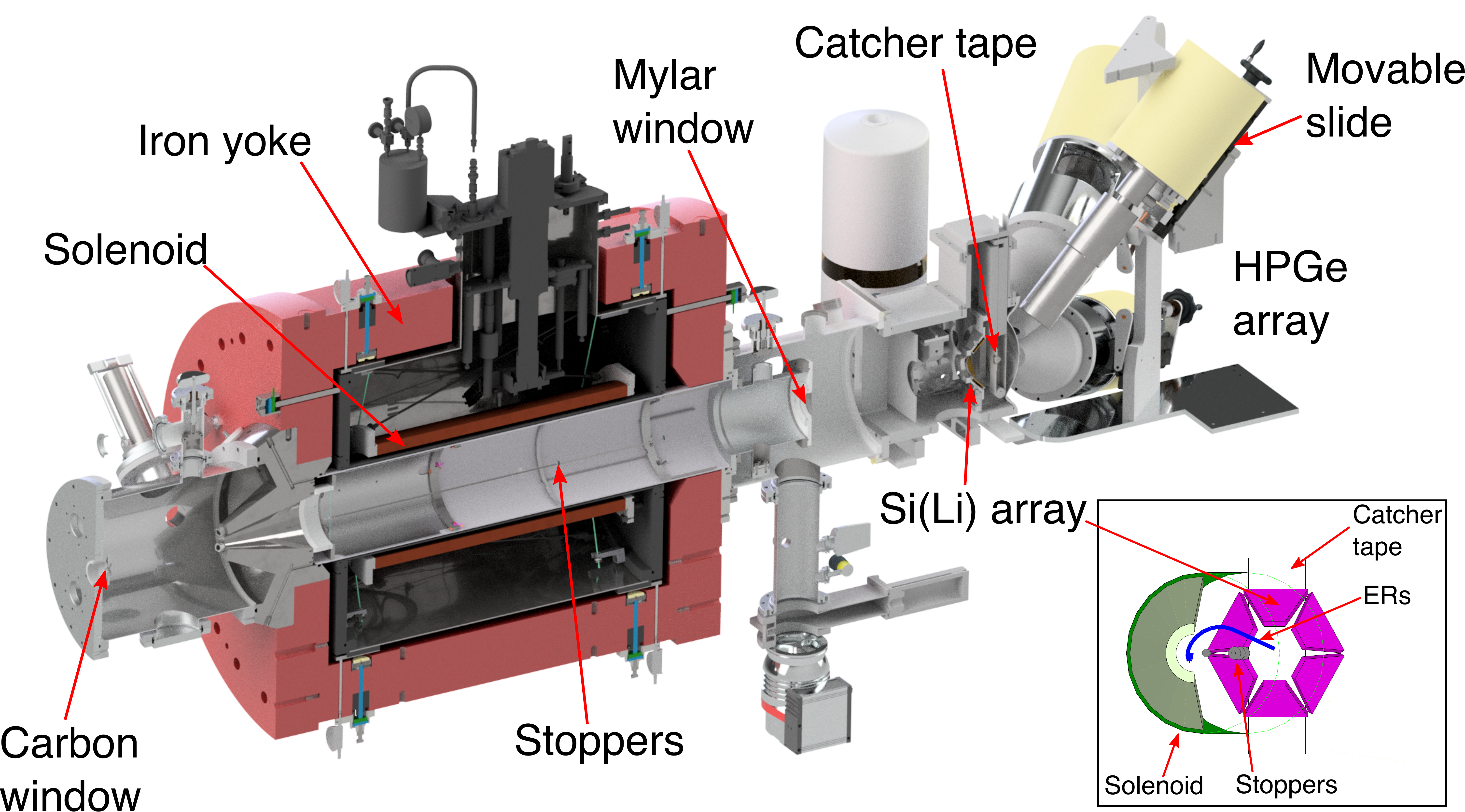}
\end{center}
\vspace*{-13pt}
\caption{Cross-sectional rendering of Solenogam and SOLITAIRE.  The inset shows a simplified model produced using the Geant4 simulation with the helical path of the evaporation residues (ERs) through the solenoid bore shown in blue.}
\label{fig:SOLENOGAM}
\end{figure*}

Solenogam (see Figure~\ref{fig:SOLENOGAM}) is a spectrometer developed at the Australian National University~(ANU), and located at the Heavy Ion Accelerator Facility~(HIAF), for the purpose of characterising the decay of long-lived nuclear states such as isomers and radioactive ground states.  The system combines a high-efficiency, high-resolution array of $\gamma$-ray and conversion-electron detectors, Solenogam, with a superconducting, solenoidal recoil separator, SOLITAIRE~\cite{Rodriguez2010}.

The solenoid allows nuclei of interest to be separated from elastically scattered beam particles and other reaction products, allowing their decay to be observed in a low-background environment, removed in space from the target location and in time from the beam burst.  With a transmission efficiency of up to 80\% and a flight time of less than a microsecond, SOLITAIRE is able to isolate states that could not be accessed using longer separators, such as the FMA~\cite{Davids1992a} and RITU~\cite{Leino1995}, albeit with a comparative loss of resolution and background reduction.  Once the evaporation residues are transported to the focus of Solenogam, their decay is observed with both electron and $\gamma$-ray detectors.  This combination allows conversion coefficients to be extracted from a simple intensity ratio so that transition multipolarities can be assigned and mixing ratios measured.

The addition of the Solenogam system to the ANU-HIAF expands on the existing capacity for conversion-electron spectroscopy and complements the existing Super-e spectrometer~\cite{Kibedi1990}.  Super-e is capable of precision measurements of conversion electrons and electron--positron pairs, however lacks extensive $\gamma$-ray capabilities.

In the following sections, descriptions of SOLITAIRE and Solenogam are presented, followed by a summary of the experiments made during the development of the system.  Lastly, the system's performance is discussed, along with plans for potential improvements and future areas of research.

%---------------------------------------------------------------------------------------
\section{SOLITAIRE}

SOLITAIRE (\underline{SOL}enoid for \underline{I}n-beam \underline{T}ransport \underline{A}nd \underline{I}dentification of \underline{R}ecoiling \underline{E}vaporation-residues)~\cite{Rodriguez2010} is a gas-filled, magnetic solenoid that separates fusion-evaporation residues from elastically scattered beam particles and transports them to the focal point of the Solenogam array (see Figure~\ref{fig:SOLENOGAM}).  Particles moving through the bore of the solenoid undergo a series of charge-changing collisions with the gas, quickly reaching an equilibrium charge-state distribution~\cite{Betz1972}.  The solenoid then acts as a converging lens with a focal length of

\begin{equation}
f \approx 4 \frac{p^{2}}{q^{2}} \frac{1}{\overline{B_{z}^{2}}L},
\end{equation}
where $p$ is the momentum of the particle, $q$ is the charge state of the ion, $L$ is the length of the solenoid, and $\overline{B_{z}^{2}}$ is the average of the squared axial field~\cite{Rodriguez2010}.  The magnetic field can be tuned such that evaporation residues of interest are transported through the solenoid while scattered beam particles and other reaction products are stopped by a series of blocking discs placed along the axis.  Initial measurements with the system have obtained a transmission of up to 80-90\%~\cite{Rodriguez2010}.  An iron return yoke ensures that the magnetic field is negligible outside the bore of the solenoid.

SOLITAIRE can be used for the precise measurement of fusion cross-sections, where knowledge of the evaporation residue charge states allows position sensitive detectors at the focal plane to reconstruct the angular distribution of evaporation residues.  Measurements of this type require a thin target so that the evaporation residues produced can escape the target with minimal loss of energy.  Alternatively, nuclear structure measurements with the Solenogam array use $\gamma$-ray and electron detectors to study the decay schemes of long-lived nuclear states.  For these measurements a thicker target can be used to maximise yield, as precise knowledge of the scattering angles is not important.  For a more detailed discussion of SOLITAIRE and its capabilities, see Ref.~\cite{Rodriguez2010}.

\begin{figure}[!htb]
\begin{center}
\includegraphics[width=\columnwidth]{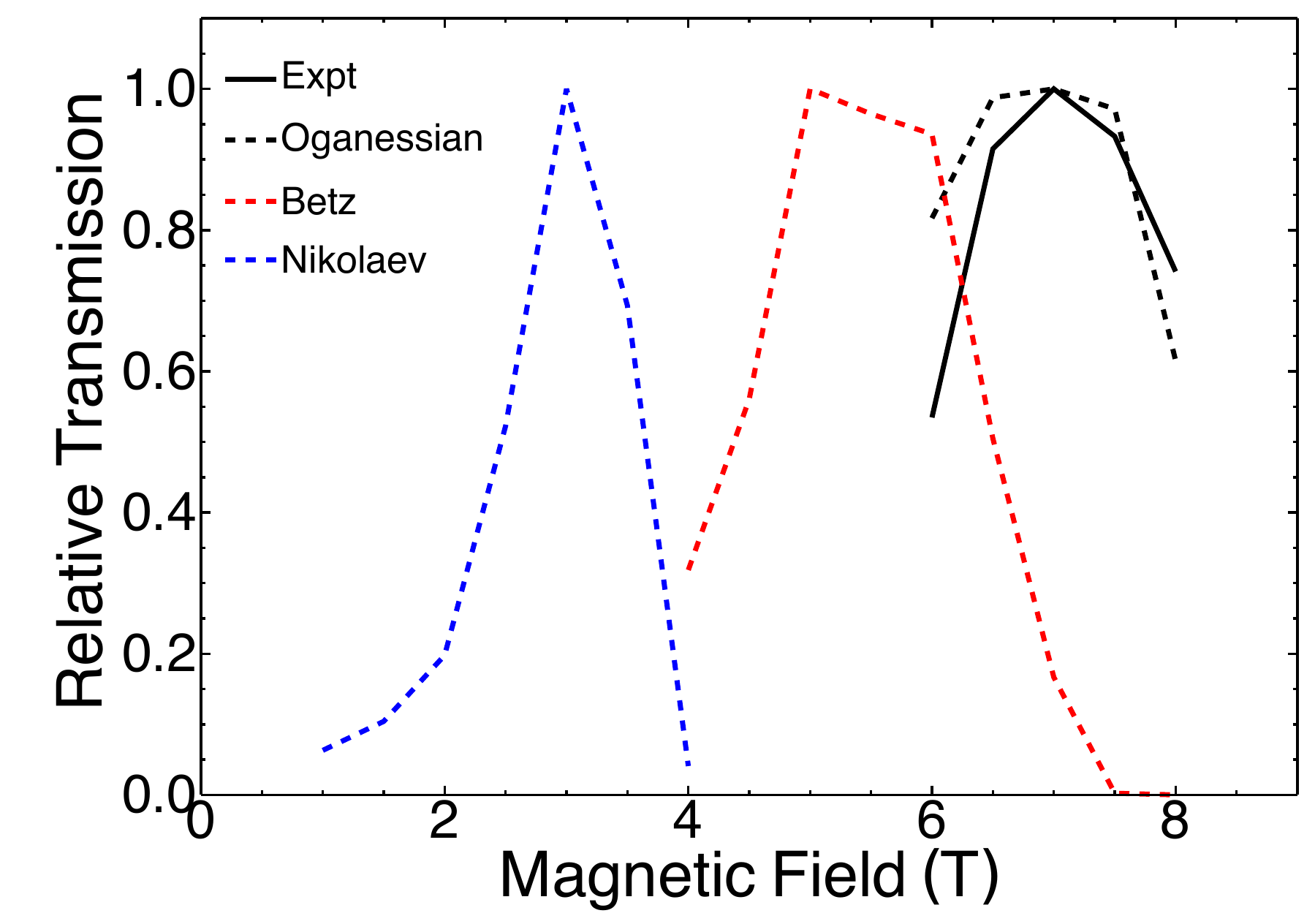}
\end{center}
\vspace*{-13pt}
\caption{Relative transmission of $^{190}$Au following the $^{176}$Yb($^{19}$F,5n) reaction with a 95~MeV $^{19}$F beam.  The dashed lines show the Simsol~\cite{Brown2011} calculations using the approaches of Betz~\cite{Betz1972}, Oganessian~\cite{Oganessian1991} and Nikolaev~\cite{Nikolaev1968} to calculate the average charge states.  The solid line is from experiment.}
\label{fig:190Pt-Trans}
\end{figure}

\begin{figure}[!htb]
\begin{center}
\includegraphics[width=\columnwidth]{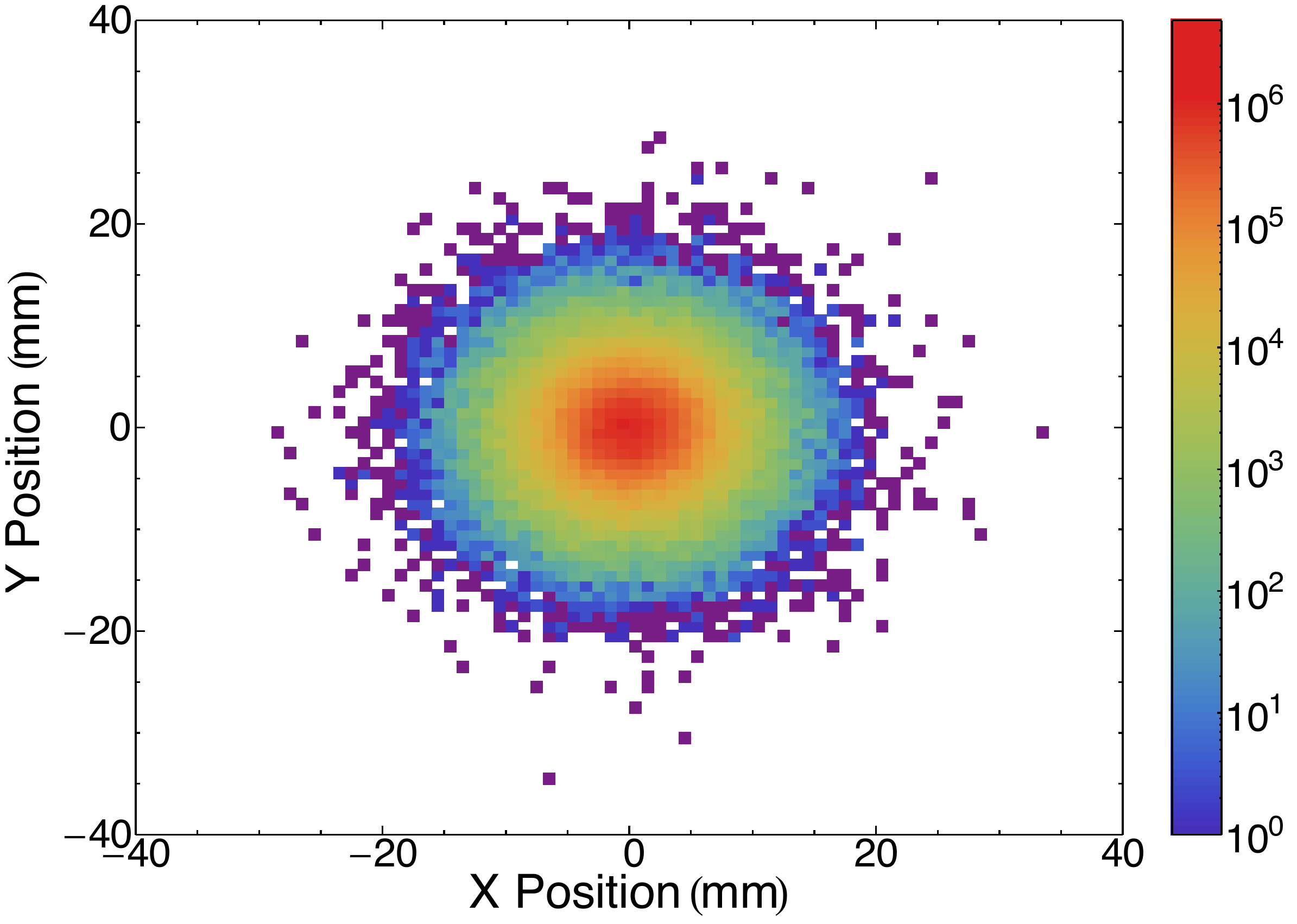}
\end{center}
\vspace*{-13pt}
\caption{Simulated distribution of $^{190}$Au implantations on the catcher tape following the $^{176}$Yb($^{19}$F,5n)$^{190}$Au reaction at 95~MeV with a magnetic field of 7~T and 0.26~Torr of He.}
\label{fig:190Pt-BeamSpot}
\end{figure}

The simulation software Simsol~\cite{Brown2011} (developed at the Australian National University) can be used to simulate the path of heavy ions through the solenoid and provides an estimate of the magnetic field required for an experiment.  Ref.~\cite{Rodriguez2010} presents a comparison of experimental results with the predictions of Simsol, giving confidence in the validity of its output.  For tuning the solenoid, the angular distribution of evaporation residues leaving the target can be determined using SRIM~\cite{Ziegler2010} and the path through the solenoid can then be simulated at different magnetic field strengths, using a number of approaches to calculate the average charge state.  Figure~\ref{fig:190Pt-Trans} shows the normalised transmission from these simulations for the $^{176}$Yb($^{19}$F,5n)$^{190}$Au reaction with a 95~MeV $^{19}$F beam, compared with measurements conducted in the current work (discussed below).  The results of the simulations and measurements were in good agreement when the approach of Oganessian et al.~\cite{Oganessian1991} was used to calculate the average charge state.  Figure~\ref{fig:190Pt-BeamSpot} shows the simulated distribution of implantations on the catcher tape predicting a beam spot with radius $\approx$1.5~cm.

The Solenogam system was originally used coupled to a solenoid with a maximum field of 6.5~T and initial results have been published (see Ref.~\cite{Dracoulis2009}).  It has since been upgraded by coupling to an 8~T solenoid, providing greater focusing power and access to a wider range of reaction possibilities, while also freeing the 6.5~T solenoid for dedicated use with radioactive-ion beams (see Ref.~\cite{Rafiei2011}).

%---------------------------------------------------------------------------------------
\section{SOLENOGAM}
\label{sec:Solenogam}

Solenogam is an array of six Si(Li) and up to seven HPGe/LEPS detectors situated at the exit of SOLITAIRE (see Figure~\ref{fig:SOLENOGAM}).  Evaporation residues transported by the solenoid are brought to a focus at the centre of the array where they are implanted in a mylar catcher tape and their subsequent decays are observed.  The array was designed with a particular goal of assigning transition multipolarities using conversion coefficients.  Due to the simultaneous measurement of both $\gamma$ rays and conversion electrons, conversion coefficients can be extracted from singles data through a direct comparison of intensities from a $\gamma$-ray and the associated conversion-electron peaks.  From $\gamma$-$\gamma$ and $\gamma$-e$^{-}$ coincidence data, intensities can be obtained from a common $\gamma$-ray gate in both the $\gamma$-$\gamma$ and $\gamma$-e$^{-}$ matrices.  This allows complex decay schemes to be resolved through the removal of contaminating lines by careful selection of $\gamma$-ray gates.  In both cases, conversion coefficients are obtained from the ratio of the efficiency-corrected intensities.

\begin{figure}[!htb]
\begin{center}
\includegraphics[width=\columnwidth]{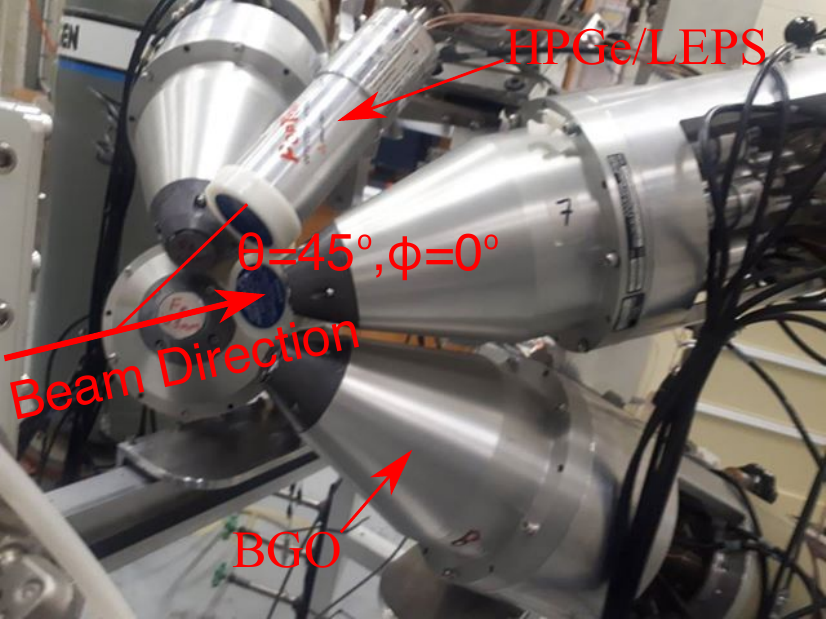}
\end{center}
\vspace*{-13pt}
\caption{The Solenogam HPGe/LEPS array consists of seven detectors, one at $\theta$=0$^{\circ}$ and six at $\theta$=45$^{\circ}$.  Note that the detector at $\phi$=180$^{\circ}$, $\theta$=45$^{\circ}$ is not mounted in this image and the array is pulled back downstream of its usual position.}
\label{fig:Solenogam-HPGeArray}
\end{figure}

The seven $\gamma$-ray detectors in Solenogam are located downstream of the focal point.  One of these detectors is located directly on the beam axis with the remaining six placed at $\theta$=45$^{\circ}$ to the beam axis with $\phi$ angles of 0$^{\circ}$, $\pm$60$^{\circ}$, $\pm$120$^{\circ}$, and 180$^{\circ}$ (see Figure~\ref{fig:Solenogam-HPGeArray}).  The front faces of the HPGe detectors are typically located about $\approx$150~mm from the focal point of the array, but are capable of being moved in and out.  The detectors can be placed inside BGO shields to provide active Compton suppression.  The improvement in spectrum quality from Compton suppression is offset by a loss of statistics, such that the choice of whether or not to use the Compton shields is best made on a case-by-case basis.

\begin{figure}[!htb]
\begin{center}
\includegraphics[width=\columnwidth]{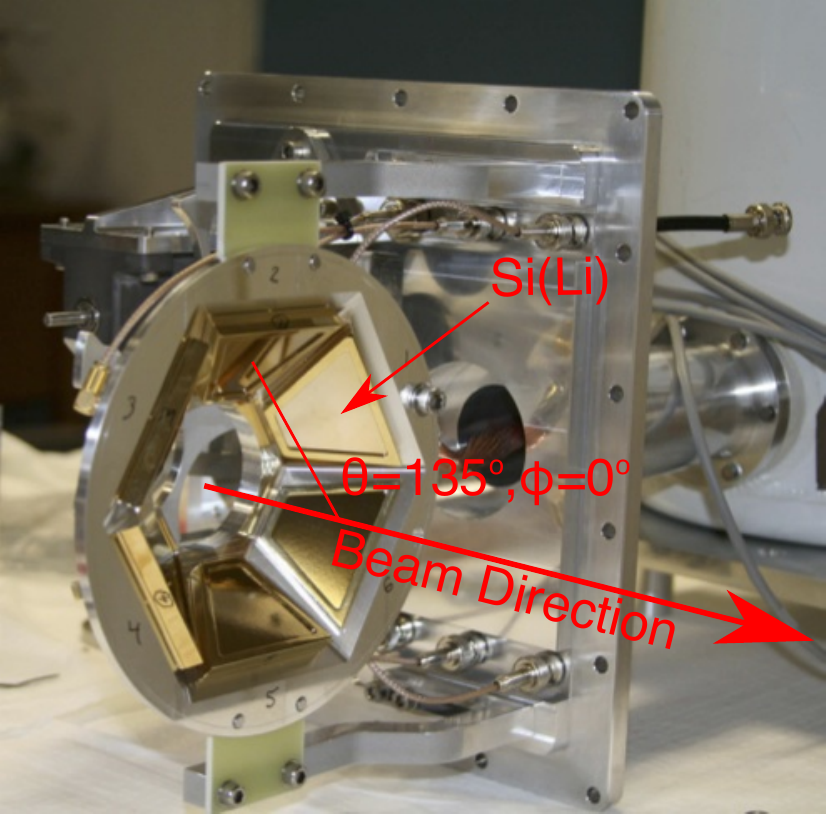}
\end{center}
\vspace*{-13pt}
\caption{The Solenogam Si(Li) array consists of six detectors at $\theta$=135$^{\circ}$.}
\label{fig:Solenogam-SiLiArray}
\end{figure}

The Si(Li) array in Solenogam is located upstream of the catcher tape and consists of six, 2-mm thick Si(Li) detectors.  The cross section of each detector is an isosceles trapezoid (see Figure \ref{fig:Solenogam-SiLiArray}) with height, major base, and minor base of 35~mm, 52~mm, and 17.5~mm, respectively, giving a total active area of 1216~mm$^{2}$ per detector and an angular coverage of $\approx$16\% of 4$\pi$ for the full Si(Li) array.  The face of each detector is covered by a 200~\si{\angstrom} gold layer.  The geometric centres of the Si(Li) detectors are $\approx$60~mm from the focal point and located at an angle of $\theta$=135$^{\circ}$ to the beam axis with $\phi$ angles that mirror the HPGe array ($\phi$=$0^{\circ}$, $\pm60^{\circ}$, $\pm120^{\circ}$, and $180^{\circ}$).

Using an open $^{152}$Eu source, the energy resolution of the detectors has been measured as $\approx$4.5~keV for the 1408-K line, slightly worse than the manufacturer-specified value of 3.5~keV.  Further loss of energy resolution can be introduced by electron straggling, caused by the implantation of the evaporation residues into the mylar catcher tape.  Both experiment and modelling have been used to evaluate this effect on the energy resolution~\cite{Whichello2014}.  The $^{152}$Eu source was placed in front of the Si(Li) array and increasingly thick pieces of mylar were placed between the source and detectors to simulate the straggling effects of the catcher tape.  These results showed that for an implantation depth of $\approx$10~$\mu$m, the conversion electrons lose $\approx$8~keV of energy leaving the tape and the measured resolution of the 1408~keV, K-Line is reduced to $\approx$9.5~keV.  As this depth can be exceeded for the products of mass-symmetric reactions, an optional degrader foil can be used to slow the evaporation residues before implantation, reducing the implantation depth and therefore the straggling.

\begin{figure}[!htb]
\begin{center}
\includegraphics[width=\columnwidth]{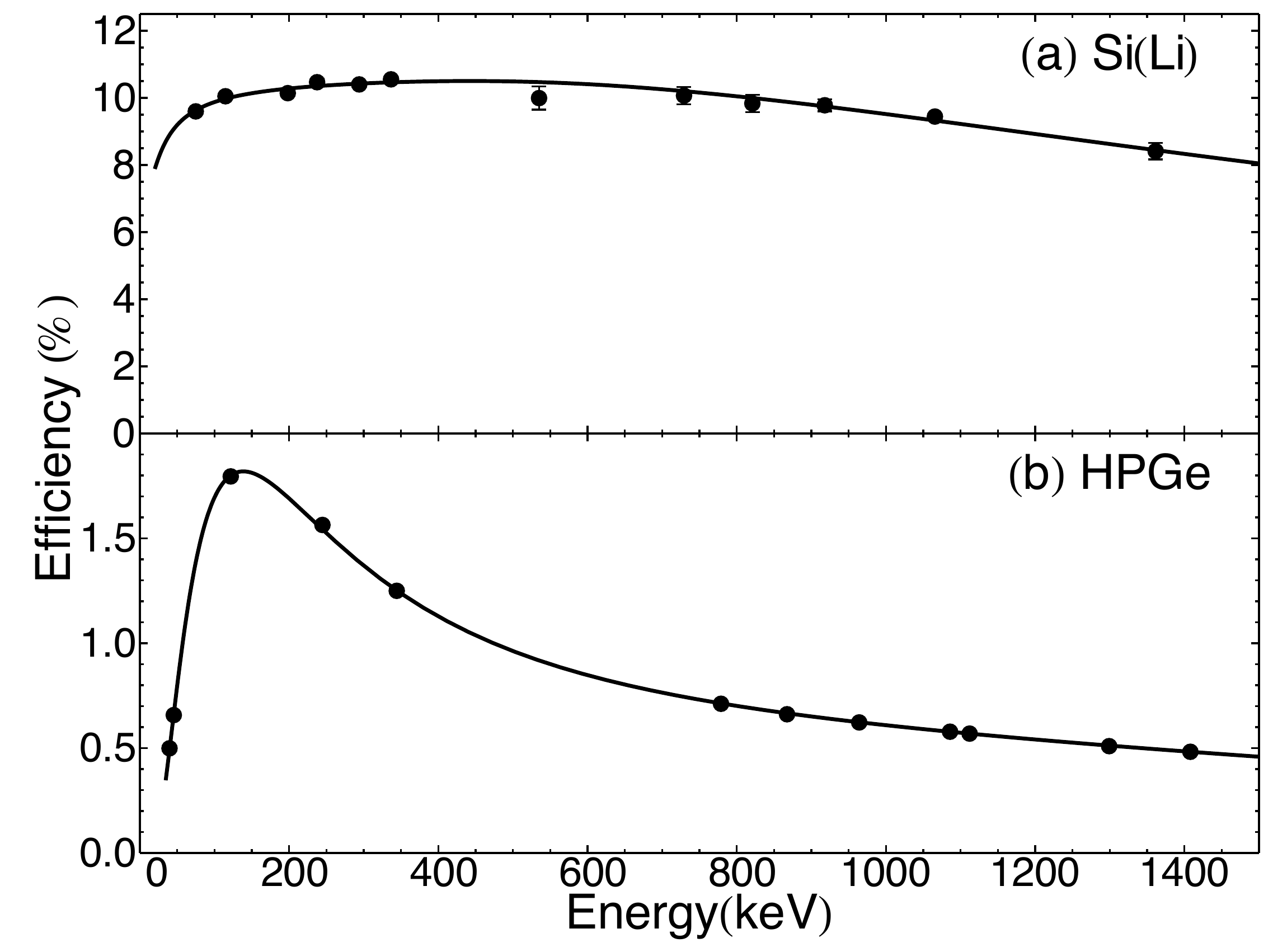}
\end{center}
\vspace*{-13pt}
\caption{Total absolute efficiency of (a)~the six element Si(Li) detector array and (b)~the HPGe detector array (five Ortec Gamma-X HPGe detectors).  Measured using an open $^{152}$Eu source.  Error bars are, in some cases, smaller than the data points.}
\label{fig:Solenogam-efficiency}
\end{figure}

Energy and efficiency calibrations can be obtained for both HPGe and Si(Li) arrays using an open $^{152}$Eu source placed at the focus of the array.  Figure~\ref{fig:Solenogam-efficiency} presents the energy-dependent absolute efficiencies for $\gamma$-ray and electron emissions.  Considering the potential radius of the beam spot after the solenoid (see Figure~\ref{fig:190Pt-BeamSpot}), a Geant4~\cite{Agostinelli2003} simulation of the array was used to determine the effects of a distributed source on the efficiency.  The effects were generally found to be small.  For a beam spot with radius 3~cm, the efficiency of the Si(Li) array decreased by $\approx$5\% and there was a negligible effect on the efficiency of the HPGe array.  Considering the Simsol calculations predict a beam spot smaller than this, in general the point source efficiencies can be used.

%---------------------------------------------------------------------------------------
\section{Commissioning experiments}

A series of experiments was performed to investigate and benchmark the performance of the system.  These were conducted at the ANU-HIAF using the 14UD tandem accelerator.  The statistical model code PACE2~\cite{Gavron1980} was used to select reactions and beam energies such that the nucleus of interest was the dominant production channel.  This is possible since, for the reactions used, neutron evaporation dominates and the thin targets and low energy loss mean that a particular xn channel is dominant.  A summary of these experiments is presented below.

\subsection[189Pb]{$^{189}$Pb}

A 32-$\mu$s isomer in $^{189}$Pb was studied using the Solenogam array, with results reported in Ref.~\cite{Dracoulis2009}. Evaporation residues from the $^{164}$Er($^{29}$Si,4n)$^{189}$Pb reaction were transported to the detector array using the 6.5~T solenoid with a 6~T magnetic field and 0.2~Torr of He gas.  Gamma-ray and electron singles were measured using one unsuppressed HPGe detector and the full Si(Li) array.  The spin and parity of the isomer was confirmed as 31/2$^{-}$ by using conversion coefficients to assign transition multipolarities, in particular the E3 and M2 isomeric decays.  This confirmation enabled the isomer's subsequent interpretation as a shears-mode bandhead~\cite{Dracoulis2009}.

%---------------------------------------------------------------------------------------
\subsection[182W]{$^{182}$W}

\begin{figure}[!htb]
\begin{center}
\includegraphics[width=\columnwidth]{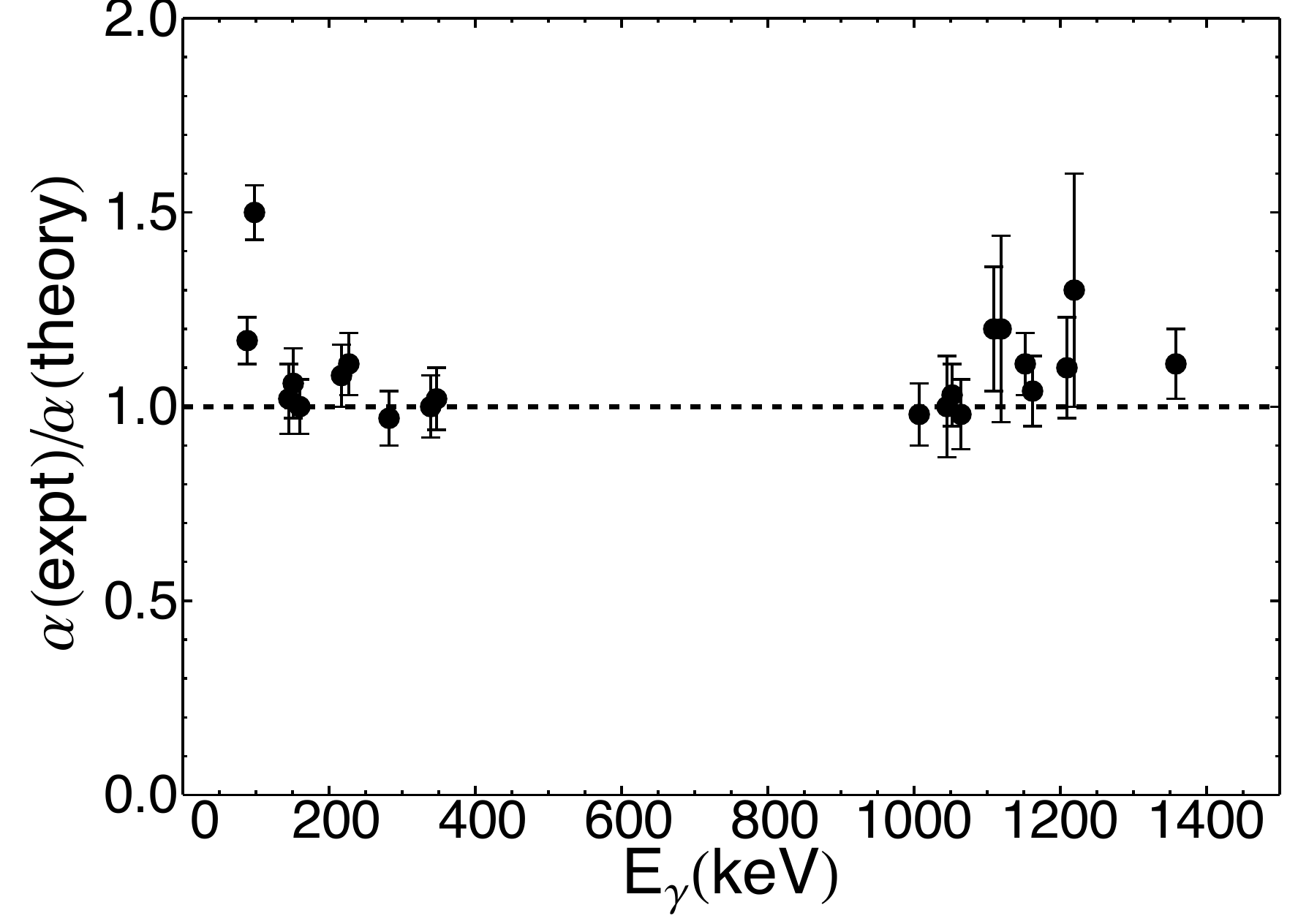}
\end{center}
\vspace*{-13pt}
\caption{Comparison of measured and theoretical conversion coefficients for $^{182}$W.  The dashed line indicates agreement between the fitted and theoretical values.}
\label{fig:182W-ICCs}
\end{figure}

An offline measurement was made using one suppressed and two unsuppressed HPGe detectors with the full electron array in order to test the ability to extract conversion coefficients from $\gamma$-e$^{-}$ coincidences (see Ref.~\cite{Gerathy2016}).  Radioactive sources of $^{182}$Re were created using the $^{176}$Yb($^{11}$B,5n) reaction with 60~MeV $^{11}$B ions incident on a 1-mg/cm$^{2}$ enriched target.  These sources were placed at the focus of the array and $\gamma$-$\gamma$ and $\gamma$-e$^{-}$ coincidence data were collected for approximately four half-lives of the $t_{1/2}$=64~hr, $\beta$ decay into $^{182}$W.  Conversion coefficients were measured for 21 transitions in the known $^{182}$W level scheme~\cite{Singh2015} by extracting the ratio of intensities in $\gamma$-gated $\gamma$-ray and conversion-electron spectra.  In general, good agreement was found with existing literature values from Ref.~\cite{Singh2015} and with theoretical values calculated using the BrIcc code~\cite{Kibedi2008} (see Figure~\ref{fig:182W-ICCs}).  A lack of agreement was seen at low energies and this has been attributed to contamination of the e$^{-}$ spectra by low-energy $\gamma$-rays.  Neglecting these two points, the weighted average is 1.03~(9).

\begin{figure}[!htb]
\begin{center}
\includegraphics[width=\columnwidth]{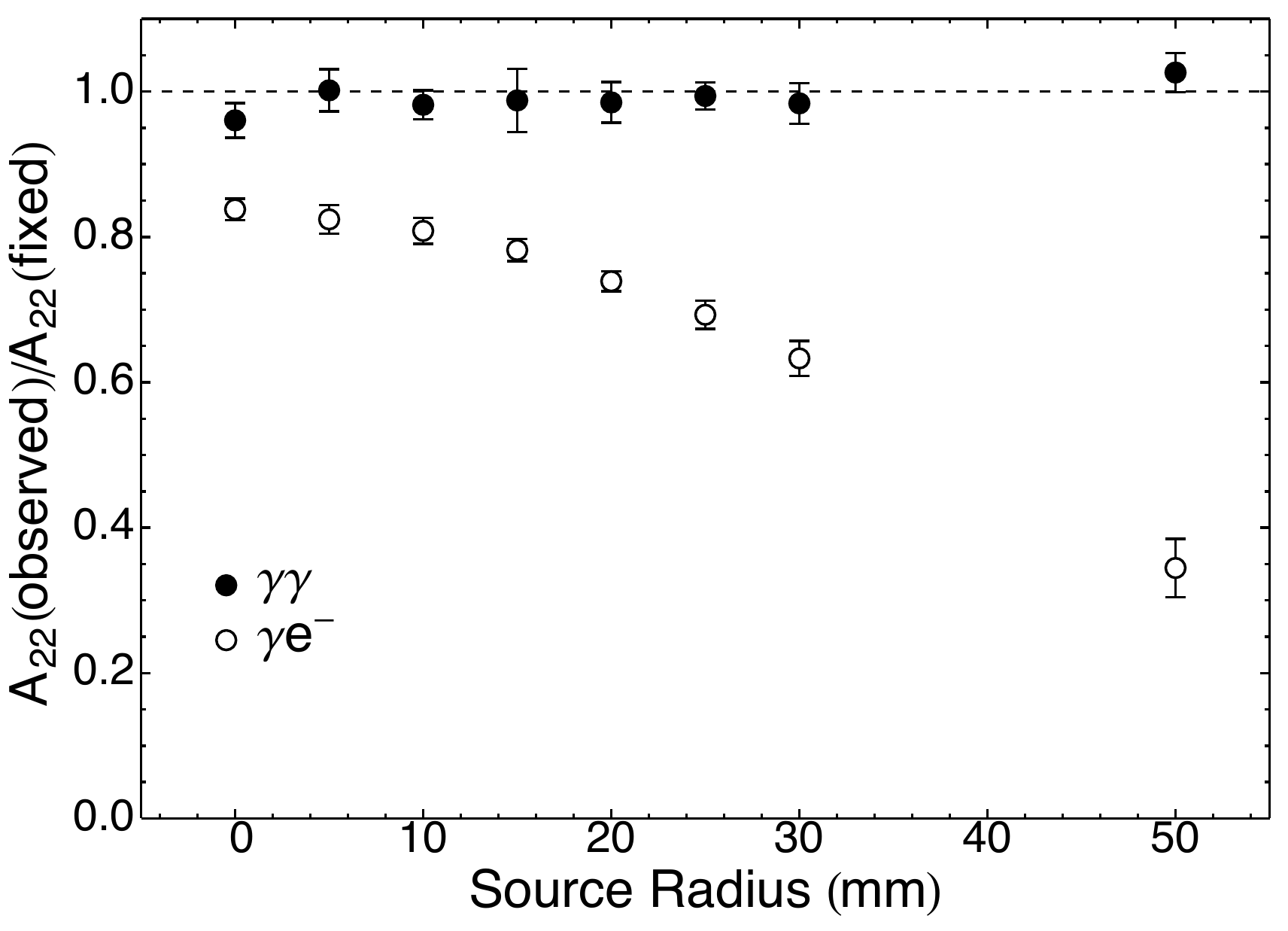}
\end{center}
\vspace*{-13pt}
\caption{Results of Geant4 simulations comparing the ratio of the anisotropy observed by the simulated detectors to the known (fixed) anisotropy for $\gamma$-$\gamma$ and $\gamma$-e$^{-}$ cascades.  The dashed line at one indicates $Q_{2}$=1 and is not a fit to the $\gamma$-$\gamma$ data.  Error bars are statistical uncertainties from fitting the modelled correlation data.}
\label{fig:G4-Qks}
\end{figure}

With the initial focus on conversion coefficients, limited consideration was given to the effects of $\gamma$-e$^{-}$ angular correlations when designing the Solenogam system.  This is a potential problem as these angular correlations could significantly affect the measured intensities in $\gamma$-ray gated spectra due to the high symmetry of the system.  As part of this work, the effects of $\gamma$-e$^{-}$ angular correlations in Solenogam were investigated to determine if they would impact the measurement of conversion coefficients in $\gamma$-ray gates and to determine the feasibility of using $\gamma$-e$^{-}$ correlations as a measurement tool.  Measured angular correlations were fitted with the formula;

\begin{equation}
W(\theta)=A_{0}[1+Q_{2}(\text{det}_{1})Q_{2}(\text{det}_{2})A_{22}P_{2}(\text{cos}\theta)],
\end{equation}
with

\begin{equation}
A_{22}=B_{2\gamma}A_{2e^{-}},
\end{equation}
where $A_{0}$ is a scaling constant, $B_{2\gamma}$ and $A_{2e^{-}}$ are the directional distribution and orientation parameters for each transition (as defined in Ref.~\cite{Steffen1975}), the $Q_{2}$ coefficients are the solid-angle corrections for each detector~\cite{Rose1953}, $P_{2}$ is the second order Legendre polynomial, and $\theta$ is the angle between detectors.  $A_{2e^{-}}$ is related to $A_{2\gamma}$ by the particle parameter, $b_{2}$, calculated using a modified version~\cite{Dowie2019} of CATAR~\cite{Pauli1975} and HEX~\cite{Liberman1971},

\begin{equation}
\label{eqn:particleparameter}
A_{2e^{-}}=b_{2}A_{2\gamma}.
\end{equation}
The Geant4 simulation discussed above in Sect.~\ref{sec:Solenogam} was used to obtain the $Q_{k}$ coefficients for the array and to study the effects of a distributed source on the angular correlations.  A modelled cascade with a fixed value of $A_{22}$ was emitted from a uniform circular source centred at the focus of the array.  The angular correlation observed by the simulated detectors was then compared with the fixed value. Figure~\ref{fig:G4-Qks} shows these results as a function of source radius.  Increasing the size of the source had little effect on the $\gamma$-$\gamma$ correlations, however, the $\gamma$-e$^{-}$ correlations were significantly attenuated for $r>1$~cm.  As the typical size of the ANU 14UD beam spot, and thus the size of the source, is $<3$~mm, the point source values of $Q_{2}$(HPGe)=0.98~(1) and $Q_{2}$(SiLi)=0.86~(2) were used for this offline measurement.  However, for on-line measurements, the attenuation from a larger beam spot will need to be considered.

\begin{figure}[!htb]
\begin{center}
\includegraphics[width=\columnwidth]{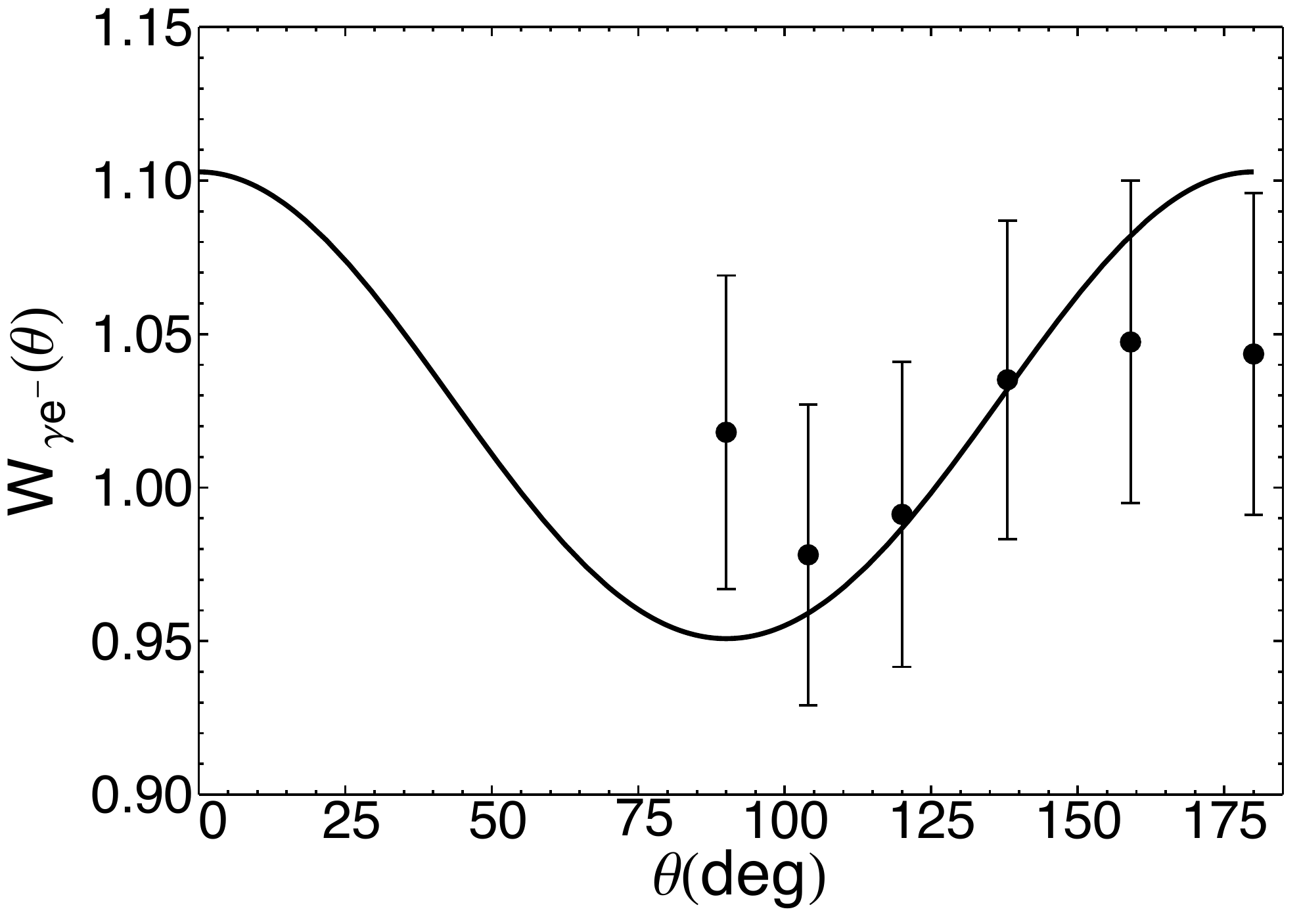}
\end{center}
\vspace*{-13pt}
\caption{Gamma-e$^{-}$ angular correlation for the $4^{+} \xrightarrow{229-\gamma} 2^{+} \xrightarrow{100-L-e^{-}} 0^{+}$ cascade in $^{182}$W. The solid line is the theoretical correlation with values of $A_{22}$=0.1 and $A_{44}$=0.002.}
\label{fig:182W-AC1}
\end{figure}

Figure~\ref{fig:182W-AC1} shows the correlations measured for the 229~keV $\gamma$-ray, 100~keV, L electron, $4^{+} \rightarrow 2^{+} \rightarrow 0^{+}$ cascade in $^{182}$W (the L-line was used as the K conversion electron was too low in energy to be detected).  Due to the small magnitude of this correlation, the effects on the measured conversion coefficients were small, $\approx$1\%.  Similar results were found for other cascades in the $^{182}$W level scheme.  For on-line measurements, the magnitude of the correlation will further decrease due to the larger source size (see Figure~\ref{fig:G4-Qks}) and hence the magnitude of these effects will be even smaller and can be largely neglected when evaluating conversion coefficients.  Furthermore, another consequence is that it is unlikely that $\gamma$-e$^{-}$ angular correlations will be a useful spectroscopic tool for Solenogam, at least with the current detector configuration.

%---------------------------------------------------------------------------------------
\subsection[184,190Pt]{$^{184,190}$Pt}

Electric monopole (E0) transitions connect states with the same spin and parity and proceed only via internal conversion.  They are sensitive to changes in the nuclear shape, making them ideal for the study of shape coexistence (see, for example, the reviews of Heyde and Wood~\cite{Heyde2011}, Wood et al.~\cite{Wood1992}, and Garrett~\cite{Garrett2016}).  For $J^{\pi} \rightarrow J^{\pi}$ transitions where $J~\neq~0$, the transition typically proceeds with a mixed M1+E2+E0 character.  The conversion coefficient for such a transition is

\begin{equation}
\label{eqn:alpha}
\alpha_{K}(\text{expt}) = \frac{\alpha_{K}(M1)+\delta^{2}(1+q_{K}^2)\alpha_{K}(E2)}{1+\delta^{2}},
\end{equation}
where $\alpha_{K}(XL)$ are the conversion coefficients for the pure M1 and E2 multipolarities, $\delta^{2}$ is the E2/M1 $\gamma$-ray mixing ratio and $q^{2}_{K}$ is the ratio of the E0 and E2 conversion electrons

\begin{equation}
q^{2}_{K}=\frac{I_{K}(E0)}{I_{K}(E2)}.
\end{equation}
In the case of a pure $0^{+} \rightarrow 0^{+}$ E0 transition, the E2 component in $q_{K}^{2}$ is taken from an E2 transition to a nearby $2^{+}$ state.

In order to fully characterise these mixed M1/E2/E0 transitions, both $\delta^{2}$ and $q^{2}_{K}$ need to be determined.  Equations of the same form as Equation~\ref{eqn:alpha} can be obtained for other electron shells (i.e $\alpha_{L}$ and $\alpha_{M}$) and this system of equations can be solved if at least two sub-shell conversion coefficients are known.  Alternatively, both $\gamma$-ray angular distributions and correlations are independent of $q^{2}_{K}$ (as E0 transitions cannot proceed via $\gamma$-ray emission) and can be used to measure $\delta$ in isolation.  A single conversion coefficient can then be used to extract $q^{2}_{K}$.  By simultaneously measuring both $\gamma$ rays and conversion electrons, Solenogam should be able to characterise E0 transitions using either of the above methods; however it should be noted that $\gamma$-ray angular distribution measurements cannot be made due to the loss of spin alignment during transit through the solenoid.

\begin{figure*}[htb]
\begin{center}
\includegraphics[width=\textwidth]{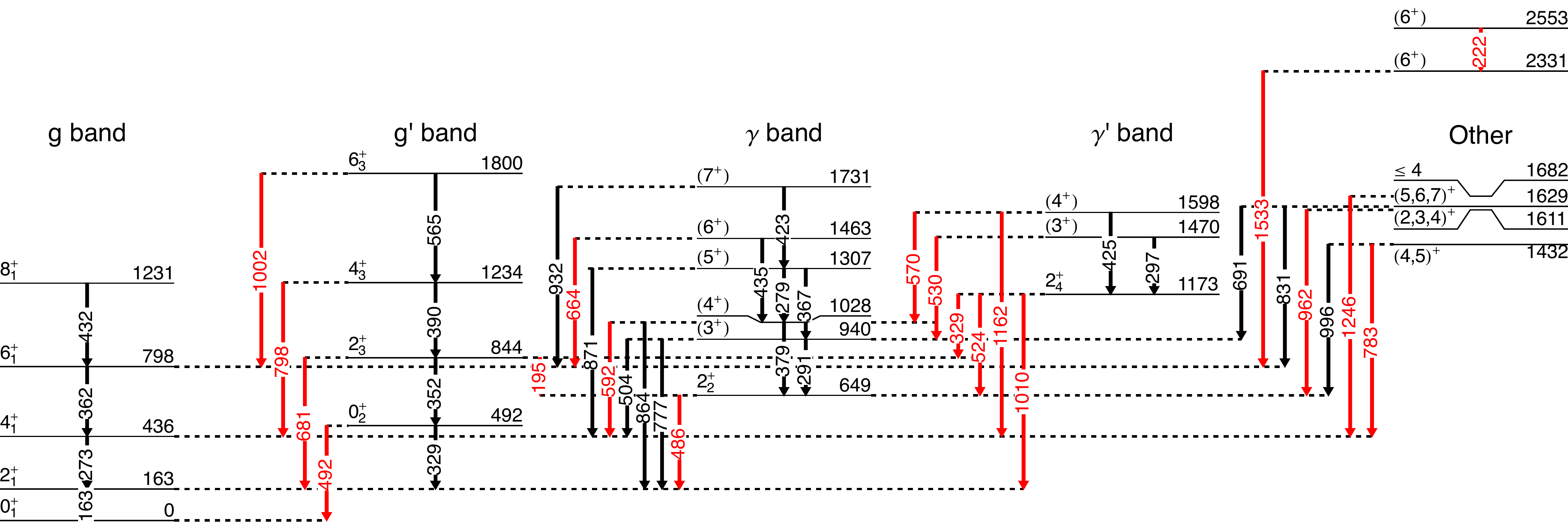}
\end{center}
\vspace*{-13pt}
\caption{Partial level scheme of $^{184}$Pt adapted from Ref.~\cite{Xu1992}.  $J^{+} \rightarrow J^{+}$ transitions with expected E0 components are highlighted in red.}
\label{fig:184Pt-levelscheme}
\end{figure*}

Previous studies have shown the presence of E0 transitions in the platinum isotopes~\cite{Xu1992}, in particular $^{184}$Pt where there are two pairs of coexisting bands.  The presence of both $K^{\pi}=0^{+}$ and $K^{\pi}=2^{+}$ bands built upon the spherical and deformed structures (see Figure~\ref{fig:184Pt-levelscheme}) provides a number of known $E0$ transitions with which to benchmark the performance of the Solenogam system.

As these were the first on-line coincidence measurements with the Solenogam system, an initial measurement of $^{190}$Pt was made.  Since $^{190}$Pt has a longer half-life ($t_{1/2}$=6.5$\times10^{11}$ years) and no daughter decay products it makes for a simple test case.  A 0.7~mg/cm$^{2}$ $^{176}$Yb target and a 95~MeV $^{19}$F beam were used to produce $^{190}$Au, with the subsequent $t_{1/2}$=43~(3)~min~\cite{Johansson1973} decay into $^{190}$Pt observed.  The beam was chopped on a long timescale of 43~min on, 43~min off (43/43~min), with data collected out of beam.  Transmission through the solenoid was measured by observing the intensity of known $^{190}$Pt $\gamma$ rays at the focus of the array throughout one measurement cycle, normalised to the beam current during the irradiation.  This was done at a range of magnetic field strengths, resulting in a maximum transmission at 7~T for 0.26~Torr He (see Figure~\ref{fig:190Pt-Trans}).  An absolute transmission through SOLTIAIRE is difficult to measure using the Solenogam array.  In Ref.~\cite{Rodriguez2010}, accurate transmissions were calculated using position sensitive particle detectors located at the focus of the solenoid and knowledge of the angular distribution of evaporation residues leaving the target.  The lack of particle detectors in the Solenogam array meant a similar approach could not be used in the current work.  An approximate transmission of $\sim$90\% was calculated based on the expected $\gamma$-ray intensities from the $^{190}$Au decay, the measured efficiency of the Solenogam array, calculated production cross-sections from PACE2, the beam current and duration of the irradiation, and the measured thickness of the target, however, it should be noted that there is significant uncertainty associated with this value.

\begin{figure*}[htb]
\centering
\includegraphics[width=0.925\textwidth]{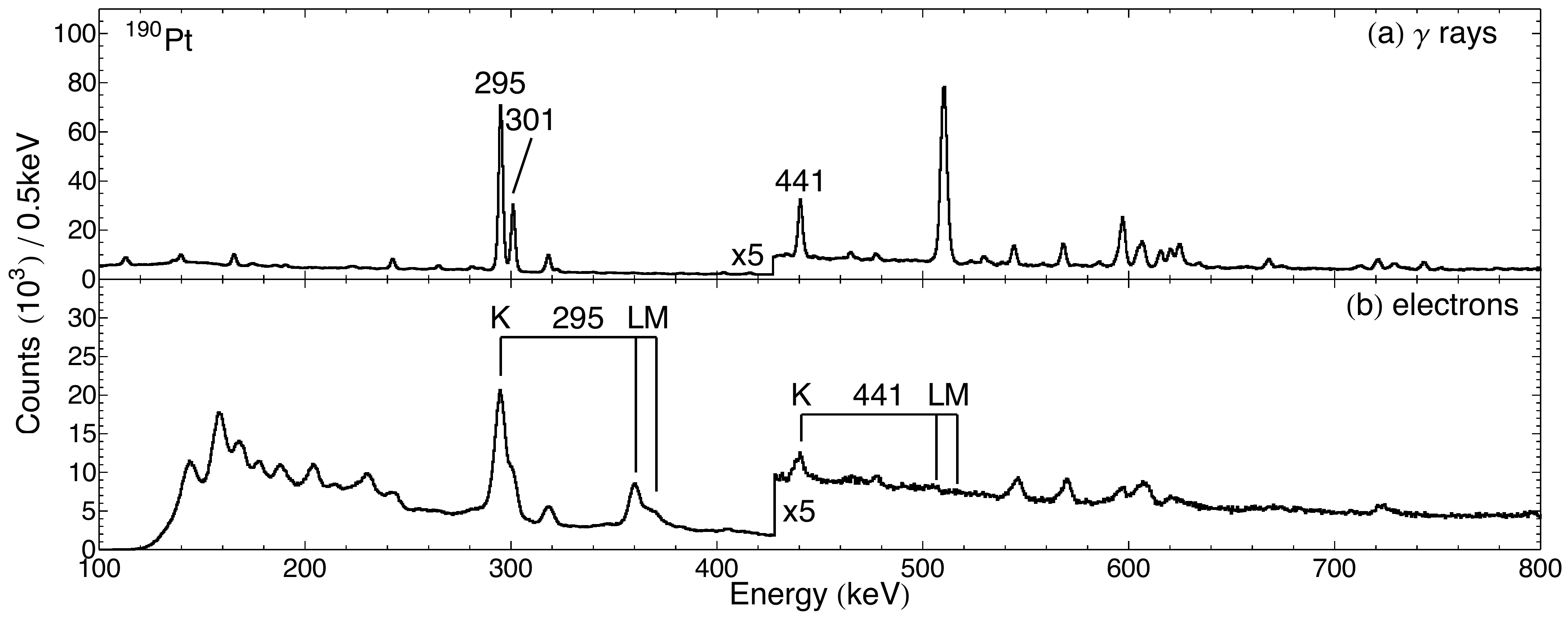}
\vspace*{-13pt}
\caption{Total projection of (a) $\gamma$ rays and (b) electrons from $\gamma$-$\gamma$ and $\gamma$-$e^{-}$ coincidence matrices from the $^{190}$Pt measurement.  Electron energies have been shifted to align the $\gamma$-ray and K-conversion peaks.}
\label{fig:190Pt-Proj}
\end{figure*}

\begin{figure*}[p]
\centering
\includegraphics[width=0.925\textwidth]{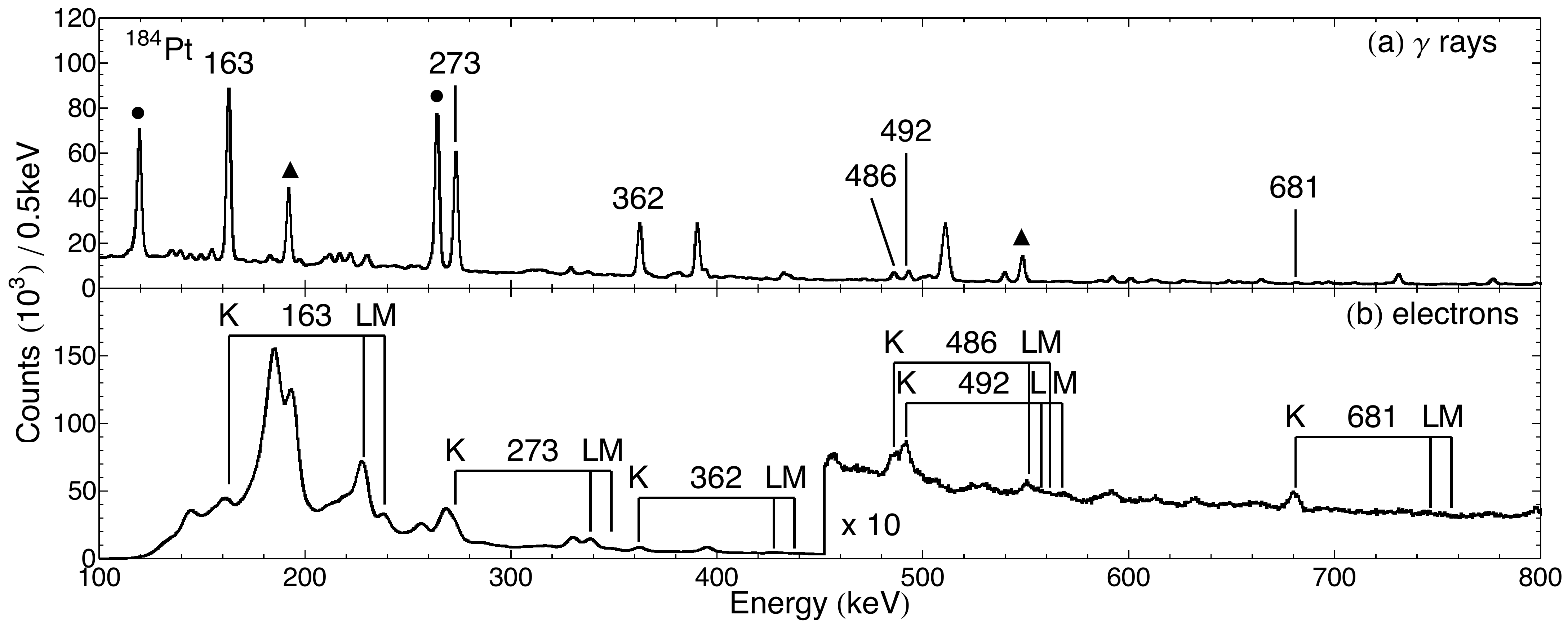}
\vspace*{-13pt}
\caption{Total projection of (a) $\gamma$ rays and (b) electrons from $\gamma$-$\gamma$ and $\gamma$-$e^{-}$ coincidence matrices from the $^{184}$Pt measurement.  Electron energies have been shifted to align the $\gamma$-ray and K-conversion peaks.  Contaminants from (\textbullet) $^{184}$Os and ($\blacktriangle$) $^{184}$Ir have been marked.}
\label{fig:184Pt-Proj}

\vspace{\floatsep}

\includegraphics[width=0.925\textwidth]{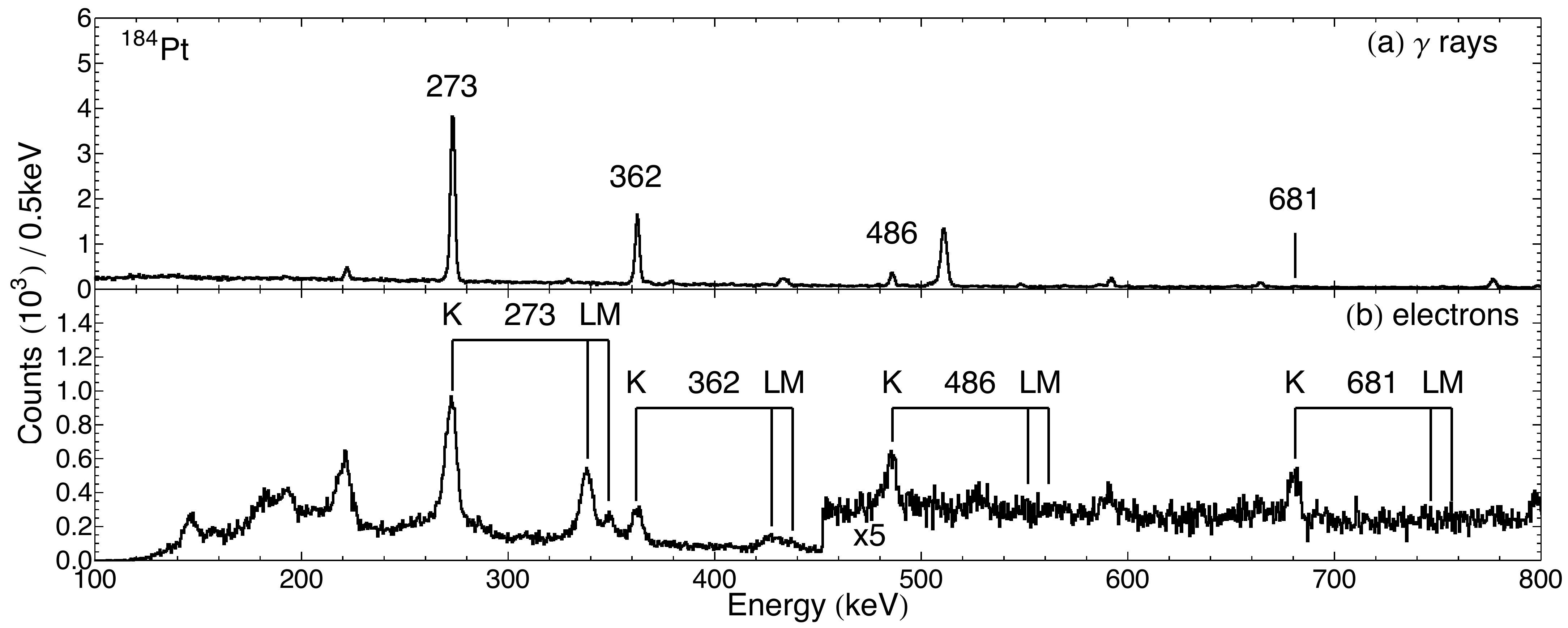}
\vspace*{-13pt}
\caption{(a) $\gamma$-ray and (b) electron coincidence spectra gated by the 163~keV, $2^{+} \rightarrow 0^{+}$ $\gamma$ ray in $^{184}$Pt.  Electron energies have been shifted to align the $\gamma$-ray and K-conversion peaks.}
\label{fig:184Pt-163Gate}

\vspace{\floatsep}

\includegraphics[width=0.925\textwidth]{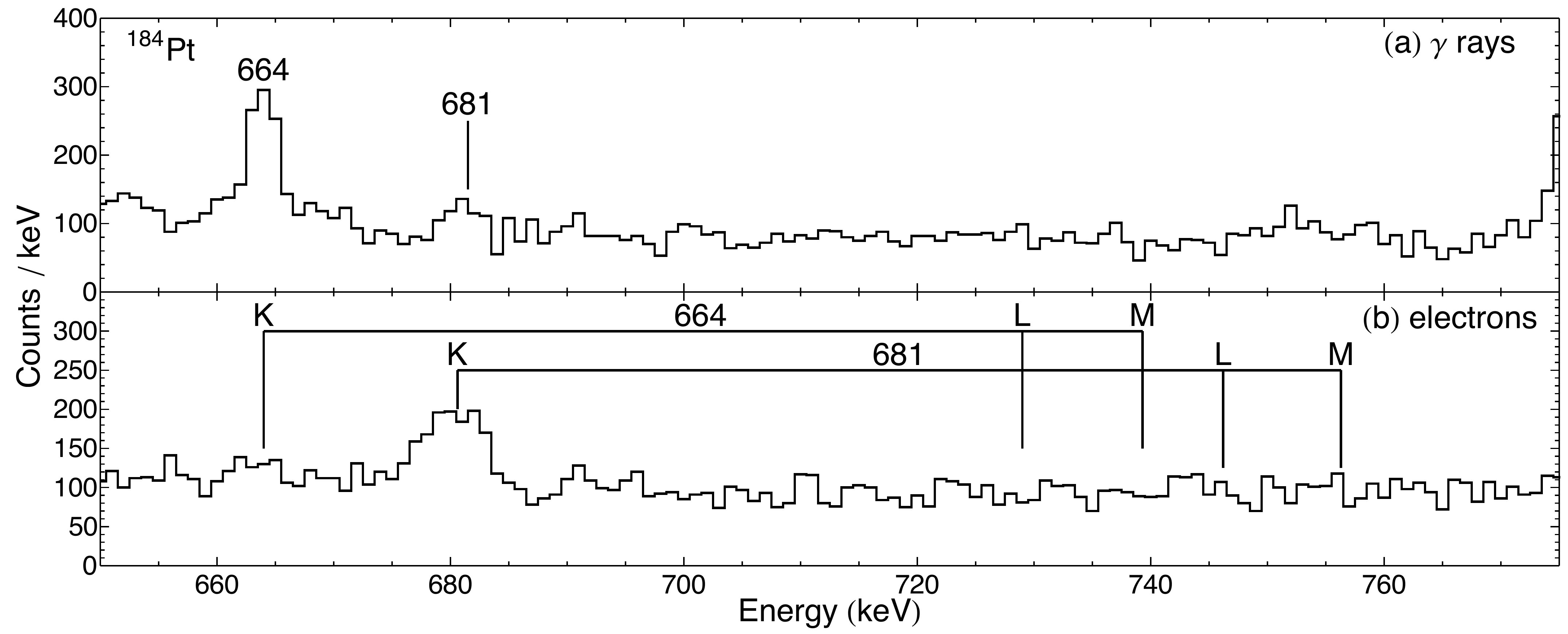}
\vspace*{-13pt}
\caption{(a) $\gamma$-ray and (b) electron coincidence spectra gated by the 163~keV, $2^{+} \rightarrow 0^{+}$ $\gamma$ ray in $^{184}$Pt, zoomed in to the region of the 681~keV transition that has a significant E0 component.  Note the comparative lack of electron intensity for the 664~keV, M1/E2 transition.  Electron energies have been shifted to align the $\gamma$-ray and K-conversion peaks.}
\label{fig:184Pt-681Spec}
\end{figure*}

In the subsequent experiment on $^{184}$Pt, the $t_{1/2}$=53~s, $\beta$ decay of $^{184}$Au was studied, produced using the $^{159}$Tb($^{30}$Si,5n)$^{184}$Au reaction with a 150~MeV $^{30}$Si beam of ions chopped at 106/106~s and incident on a 0.8 mg/cm$^{2}$ target.  Transmission through SOLITAIRE was maximised with a 7~T magnetic field for 0.25~Torr of He gas.  For both the $^{190}$Pt and $^{184}$Pt measurements, five HPGe detectors and one LEPS detector were used.  The HPGe detectors at $\phi$=$\pm$60$^{\circ}$, and $\pm$120$^{\circ}$ were placed inside BGO Compton shields.  The total projections of $\gamma$ rays and electrons from $\gamma$-$\gamma$ and $\gamma$-$e^{-}$ coincidence matrices from the $^{190}$Pt and $^{184}$Pt measurements are shown in Figures~\ref{fig:190Pt-Proj} and~\ref{fig:184Pt-Proj}, respectively.  Spectra gated by the 163~keV, $2^{+} \rightarrow 0^{+}$ $\gamma$ ray in $^{184}$Pt are shown in Figures~\ref{fig:184Pt-163Gate} and~\ref{fig:184Pt-681Spec}.

\begin{table*}[p]
\centering
\caption{Conversion coefficients measured for $^{184,190}$Pt.  Literature conversion coefficients are from Refs.~\cite{Johansson1973,Baglin2010}. Theoretical $\alpha_{K}(M1)$ and $\alpha_{K}(E2)$ values are from BrIcc~\cite{Kibedi2008}.  Multipolarity assignments are from Refs.~\cite{Baglin2010,Singh2003}.}
\label{tab:Pt-ICCs}
\begin{tabular}{c c c c c c c c c c}
\toprule
\toprule
Nucleus & $E_{\gamma}$(keV) & $E_{i} \rightarrow E_{f}$ & $J_{i} \rightarrow J_{f}$ & \multicolumn{4}{c}{$\alpha_{K}\times 100$} & Multipolarity\\
\cline{5-8}
& & & & This work & Literature & M1 & E2 & \\
\midrule
$^{190}$Pt
& 296  & 296~$\rightarrow$~0    & 2$^{+}$~$\rightarrow$~0$^{+}$      & 5.3 (4)  & 6.1 (5) & 26.4 & 6.37 & E2	     \\
& 302  & 598~$\rightarrow$~296  & 2$^{+}$~$\rightarrow$~2$^{+}$      & 5.7 (4)  & 5.9 (5) & 25.0 & 6.06 & M1/E2     \\
& 319  & 917~$\rightarrow$~598  & 3$^{+}$~$\rightarrow$~2$^{+}$      & 4.3 (7)  & 5.7 (6) & 21.5 & 5.29 & E2        \\	
& 323  & 921~$\rightarrow$~598  & 0$^{+}$~$\rightarrow$~2$^{+}$      & 7.1 (3)  & 5.0 (6) & 20.6 & 5.09 & E2	      \\
& 441  & 737~$\rightarrow$~598  & 4$^{+}$~$\rightarrow$~2$^{+}$      & 2.3 (3)  & 2.9 (4) & 8.97 & 2.42 & E2	      \\
& 531  & 1128~$\rightarrow$~598 & (4$^{+}$)~$\rightarrow$~2$^{+}$    & 1.5 (8)  & <2.6    & 5.54 & 1.61 & (E2)	      \\
& 605  & 1203~$\rightarrow$~598 & 2$^{+}$~$\rightarrow$~2$^{+}$      & 2.3 (4)  & 3.4 (5) & 3.91 & 1.21 & M1(/E2)	  \\
& 616  & 1353~$\rightarrow$~737 & 3$^{-}$~$\rightarrow$~4$^{+}$      & 0.5 (1)  & 0.44 (8)& 3.75 & 1.16 & E1		  \\	
& 621  & 917~$\rightarrow$~ 296 & 3~$\rightarrow$~2$^{+}$            & 1.0 (2)  & 1.6 (2) & 3.68 & 1.15 & M1/E2	  \\	
& 625  & 921~$\rightarrow$~ 296 & 0~$\rightarrow$~2$^{+}$            & 1.1 (2)  & 1.08 (2)& 3.62 & 1.13 & E2	      \\	
& 797  & 1395~$\rightarrow$~598 & 2$^{+}$~$\rightarrow$~2$^{+}$      & 30 (20)  & 17 (4)  & 1.93 & 0.69 & E0(/M1/E2) \\
& 907  & 1203~$\rightarrow$~296 & 2$^{+}$~$\rightarrow$~2$^{+}$      & 3.8 (5)  & 3.8 (5) & 1.39 & 0.53 & E0(/M1/E2) \\
& 1005 & 1602~$\rightarrow$~598 & (1,2)$^{+}$~$\rightarrow$~2$^{+}$  & <0.5     & -       & 1.07 & 0.44 & M1/E2	  \\
& 1099 & 1395~$\rightarrow$~296 & 2~$\rightarrow$~2$^{+}$            & 2.1 (5)  & 2.7 (5) & 0.85 & 0.37 & E0(/M1/E2) \\	
\midrule
$^{184}$Pt 
& 195  & 844 $\rightarrow$649  & 2$^{+}$~$\rightarrow$~2$^{+}$       & 123 (30) & 90 (30) & 82.6 & 18.1 & E0/M1/E2 \\
& 222  & 2553$\rightarrow$2331 & (6$^{+}$)~$\rightarrow$~(6$^{+}$)   & 91 (8)   & -       & 57.6 & 13.1 & E0/M1/E2 \\
& 273  & 436 $\rightarrow$163  & 4$^{+}$~$\rightarrow$~2$^{+}$       & 9.7 (7)  & 9.1 (20)& 32.6 & 7.73 & E2       \\
& 329  & 1173$\rightarrow$844  & 2~$\rightarrow$~2                   & 20 (5)   & 19 (5)  & 19.5 & 4.86 & M1       \\
& 362  & 798 $\rightarrow$436  & 6$^{+}$~$\rightarrow$~4$^{+}$       & 5.1 (4)  & 4.1 (7) & 15.2 & 3.86 & E2       \\
& 435  & 1463$\rightarrow$1028 & (6)$^{+}$~$\rightarrow$~(4)$^{+}$   & 14 (3)   & -       & 9.36 & 2.52 & -        \\
& 486  & 649 $\rightarrow$163  & 2$^{+}$~$\rightarrow$~2$^{+}$       & 5.7 (6)  & 5.1 (6) & 6.95 & 1.95 & M1/E2    \\
& 504  & 940 $\rightarrow$436  & (3)$^{+}$~$\rightarrow$~4$^{+}$     & 5 (1)    & -       & 6.35 & 1.80 & -        \\
& 524  & 1173$\rightarrow$649  & 2$^{+}$~$\rightarrow$~2$^{+}$       & 37 (24)  & 36 (4)  & 5.68 & 1.64 & E0/M1/E2 \\
& 530  & 1470$\rightarrow$940  & 3$^{+}$~$\rightarrow$~3$^{+}$       & 20 (13)  & >40     & 5.54 & 1.61 & E0/M1/E2 \\
& 570  & 1598$\rightarrow$1028 & 4$^{+}$~$\rightarrow$~4$^{+}$       & -        & 7.8 (16)& 4.58 & 1.37 & E0/M1/E2 \\
& 592  & 1028$\rightarrow$436  & (4)$^{+}$~$\rightarrow$~4$^{+}$     & 1.0 (4)  & 2.2(3)  & 4.15 & 1.26 & M1/E2    \\
& 664  & 1463$\rightarrow$798  & (6)$^{+}$~$\rightarrow$~6$^{+}$     & 1.3 (3) & 1.26(14) & 3.09 & 1.00 & M1/E2    \\
& 681  & 844 $\rightarrow$163  & 2$^{+}$~$\rightarrow$~2$^{+}$       & 30 (6)   & 30(3)   & 2.89 & 0.95 & E0/M1/E2 \\
& 777  & 940 $\rightarrow$163  & (3)$^{+}$~$\rightarrow$~2$^{+}$     & 0.85 (3) & 1.1(2)  & 2.06 & 0.72 & (M1/)E2  \\
& 798  & 1234$\rightarrow$436  & 4$^{+}$~$\rightarrow$~4$^{+}$       & 4.4 (9)  & 4.5(6)  & 1.93 & 0.69 & E0/M1/E2 \\
& 831  & 1629$\rightarrow$798  & (5,6,7)$^{+}$~$\rightarrow$~6$^{+}$ & 3.2 (6)  & 1.5(4)  & 1.74 & 0.63 & M1       \\
& 871  & 1307$\rightarrow$436  & (5)$^{+}$~$\rightarrow$~4$^{+}$     & 1.1 (3)  & 0.66(16)& 1.54 & 0.58 & M1/E2    \\
& 996  & 1432$\rightarrow$436  & 4$^{+}$~$\rightarrow$~4$^{+}$       & -        & <1.4    & 1.10 & 0.45 & M1/E2    \\
& 1002 & 1800$\rightarrow$798  & 6$^{+}$~$\rightarrow$~6$^{+}$       & -        & 1.1 (4) & 1.08 & 0.44 & M1/E2    \\
& 1010 & 1173$\rightarrow$163  & 2$^{+}$~$\rightarrow$~2$^{+}$       & -        & 1.2 (4) & 1.06 & 0.43 & M1       \\
& 1162 & 1598$\rightarrow$436  & 4$^{+}$~$\rightarrow$~4$^{+}$       & -        & $\le$0.6& 0.74 & 0.33 & D+Q      \\
\bottomrule
\bottomrule
\end{tabular}
\end{table*}

Internal conversion coefficients were extracted from coincidence data for 14 transitions in $^{190}$Pt and 17 in $^{184}$Pt (see Table~\ref{tab:Pt-ICCs}).  In both cases, good agreement and a similar level of precision was found when compared with the existing literature~\cite{Xu1992,Johansson1973}.  The measured conversion coefficients for many transitions were consistent with theoretical values from BrIcc~\cite{Kibedi2008} (see Table~\ref{tab:Pt-ICCs} and Figure~\ref{fig:184,190Pt-ICCs}), however, a number of $J^{\pi} \rightarrow J^{\pi}$ transitions were found to have an E0 component, as indicated by a conversion coefficient significantly greater than the $\alpha_{K}(M1)$ value.  Attempts to measure both the M1/E2 and E2/E0 mixing ratios for these transitions were unsuccessful, as described below.

Attempts to measure both $\delta^{2}$ and $q_{K}^{2}$ using conversion coefficients from different electron shells were unsuccessful due to poor electron resolution and statistics.  Figure~\ref{fig:184Pt-681Spec} shows the $\gamma$ ray and conversion electrons from the 681~keV transition in $^{184}$Pt, in spectra that are both gated by the 163~keV, $2^{+} \rightarrow 0^{+}$ transition.  While the K conversion-electron line can be clearly resolved, the corresponding L and M lines are at the level of the background.  The other $J^{\pi} \rightarrow J^{\pi}$ transitions in the nucleus had weaker conversion lines than the 681~keV transition and faced the same problem.

\begin{figure}[!htb]
\begin{center}
\includegraphics[width=\columnwidth]{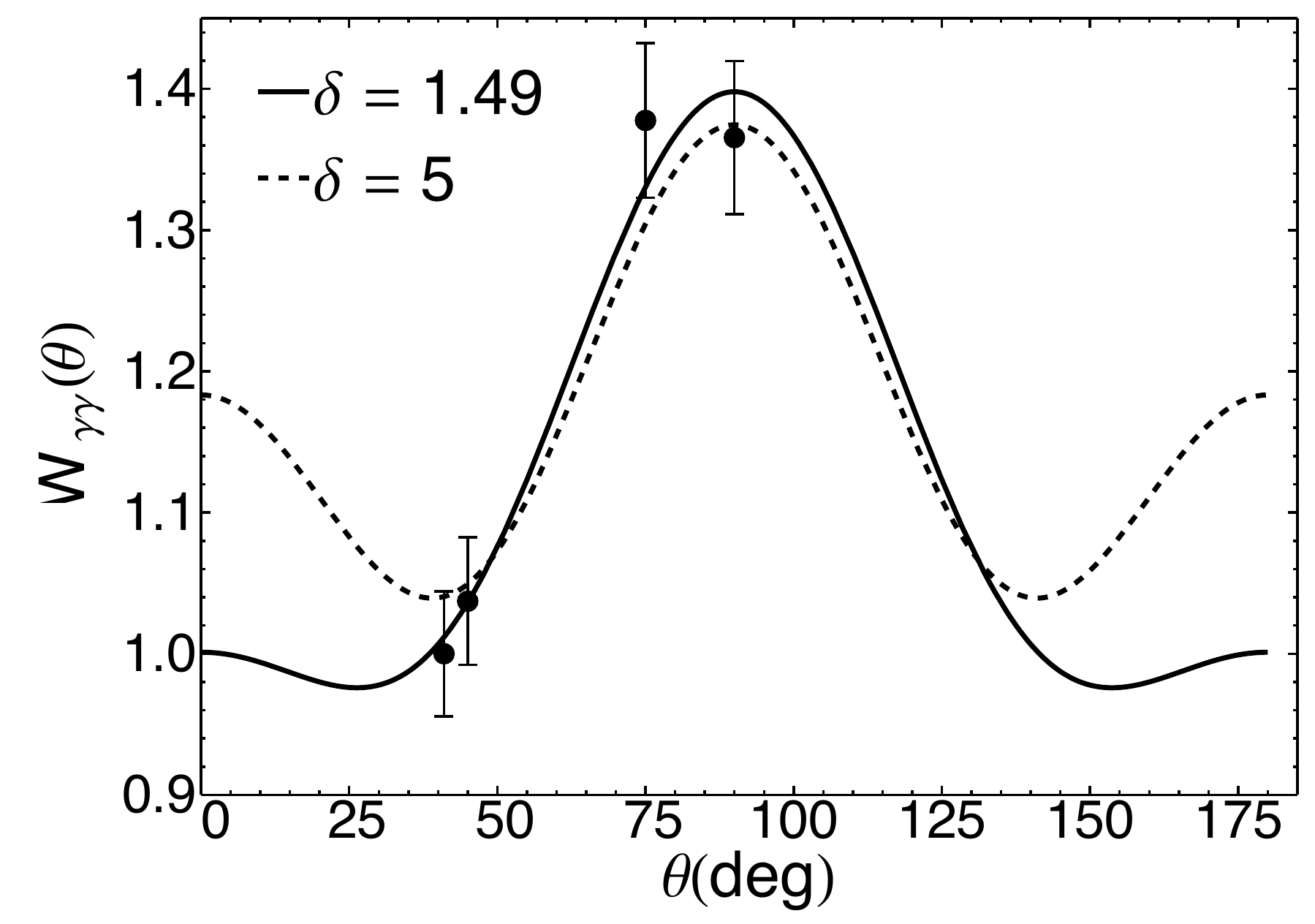}
\end{center}
\vspace*{-13pt}
\caption{Gamma--gamma angular correlation for the 486-163~keV, $2^{+}_{2} \rightarrow 2^{+}_{1} \rightarrow 0^{+}_{1}$ cascade in $^{184}$Pt.  The angular differences in the array are not sensitive to the mixing ratio, $\delta$.  Detectors that are separated by <30$^{\circ}$ (>150$^{\circ}$) are needed.}
\label{fig:184Pt-486AC}
\end{figure}

Attempts to measure $\gamma$-$\gamma$ angular correlations were similarly unsuccessful.  Figure~\ref{fig:184Pt-486AC} shows the observed correlations for the 486-163~keV, $2^{+}_{2} \rightarrow 2^{+}_{1} \rightarrow 0^{+}_{1}$ cascade along with the theoretical correlations for an M1/E2 transition with $\delta=+1.49$ and $\delta=+5$.  It is clear from the location of the experimental points that the angular separations between detectors in the present array are not sufficient to resolve such ambiguity in the M1/E2 mixing ratio.  

Without resolving these problems, the Solenogam system, on its own, is unable to collect sufficient spectroscopic data to extract the $\delta^{2}$ and $q_{K}^{2}$ values for these mixed M1/E2/E0 transitions.  However, with knowledge of the M1/E2 mixing ratio from another source, the $q_{K}^{2}$ value can be extracted using the measured $\alpha_{K}$ value.

Table~\ref{tab:E0s} contains a summary of the E0 transitions in $^{184}$Pt measured in the current work.  For the transitions where the E2/M1 mixing ratio is known, $q^{2}_{K}$ coefficients have been calculated using Equation~\ref{eqn:alpha} and the weighted average of conversion coefficients from both the current work and Xu et al.~\cite{Xu1992}.  Where the mixing ratio was not known, limits on $q^{2}_{K}$ have been placed by assuming either a pure M1 ($\delta$=0) or pure E2 ($\delta \rightarrow \infty$) transition.  The agreement between the conversion coefficients measured in the current work and those from Ref.~\cite{Xu1992}, coupled with the lack of new mixing ratios measured in the current work, means the $q^{2}_{K}$ coefficients in Table~\ref{tab:E0s} are largely unchanged from existing values~\cite{Xu1992} and are still consistent with the interpretation of the $^{184}$Pt level scheme as two pairs of coexisting bands.

\begin{figure}[!htb]
\begin{center}
\includegraphics[width=\columnwidth]{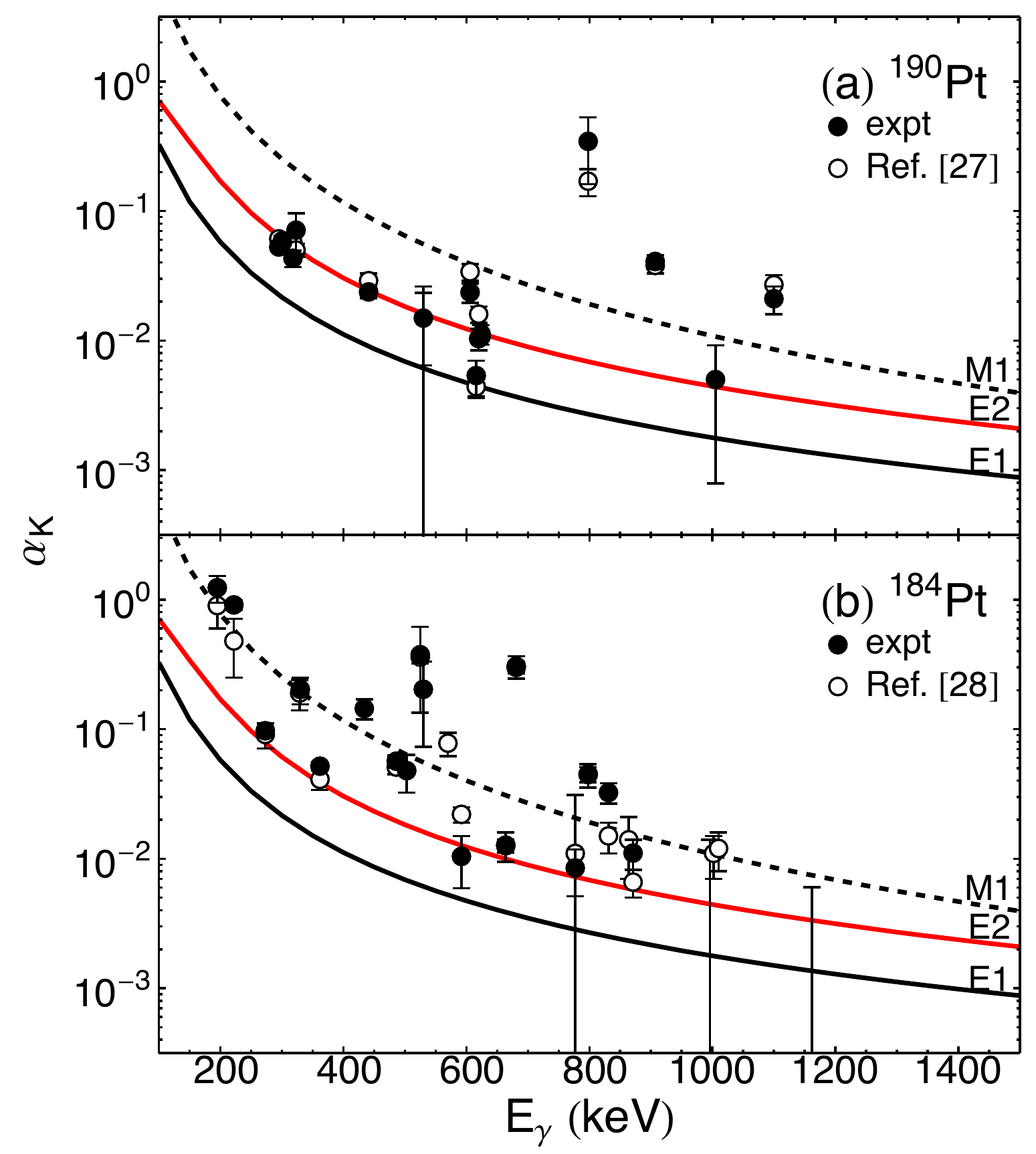}
\end{center}
\vspace*{-13pt}
\caption{Internal conversion coefficients in (a)~$^{190}$Pt and (b)~$^{184}$Pt showing both values from the current work and Refs.~\cite{Johansson1973,Baglin2010}.  The curves are theoretical values from BrIcc~\cite{Kibedi2008}.}
\label{fig:184,190Pt-ICCs}
\end{figure}

\begin{table}[!htb]
\small
\centering
\caption{Summary of the E0 transitions in $^{184}$Pt measured in the current work.  Where possible, $\alpha_{K}(expt)$ is the weighted average of the current work and Ref.~\cite{Xu1992}.  Theoretical $\alpha_{K}(M1)$ and $\alpha_{K}(E2)$ values are from BrIcc~\cite{Kibedi2008}, while values of $\delta$ are from Ref.~\cite{Xu1992a}.}
\label{tab:E0s}
\resizebox{\columnwidth}{!}{
\begin{threeparttable}
\begin{tabular}{c c c c c c c}
\toprule
\toprule
$E_{\gamma}$(keV) & $J_{i}\rightarrow J_{f}$ & \multicolumn{3}{c}{$\alpha_{K}\times 100$}& $\delta$(E2/M1) & $q^{2}_{K}$(E0/E2) \\
\cline{3-5}
& & expt & M1 & E2 &  & \\
\midrule
492  & 0$_{g'}^{+}$~$\rightarrow$~0$_{g}^{+}$           & -        & -    & -    & -                         & 8.9 (9)\tnote{a} \\
486  & 2$_{\gamma}^{+}$~$\rightarrow$~2$_{g}^{+}$       & 5.4 (4)  & 6.95 & 1.95 & +0.49 (7)                 & 0     \\
681  & 2$_{g'}^{+}$~$\rightarrow$~2$_{g}^{+}$           & 30 (3)   & 2.89 & 0.95 & -1.2$^{+0.5}_{-3.5}$      & 50$^{+17}_{-115}$\\
195  & 2$_{g'}^{+}$~$\rightarrow$~2$_{\gamma}^{+}$      & 110 (20) & 82.6 & 18.1 & -                         & >5    \\
1010 & 2$_{\gamma'}^{+}$~$\rightarrow$~2$_{g}^{+}$      & 1.2 (4)  & 1.06 & 0.43 & -                         & >1.8  \\
524  & 2$_{\gamma'}^{+}$~$\rightarrow$~2$_{\gamma}^{+}$ & 37 (4)   & 5.68 & 1.64 & -                         & >21   \\
329  & 2$_{\gamma'}^{+}$~$\rightarrow$~2$_{g'}^{+}$     & 19 (5)   & 19.6 & 4.86 & -                         & <2.9  \\
530  & 3$_{\gamma'}^{+}$~$\rightarrow$~3$_{\gamma}^{+}$ & 20 (8)   & 5.68 & 1.61 & -                         & >11   \\
592  & 4$_{\gamma}^{+}$~$\rightarrow$~4$_{g}^{+}$       & 1.4 (2)  & 4.15 & 1.26 & -2.25$^{+1.08}_{-\infty}$ & <0.11 \\
798  & 4$_{g'}^{+}$~$\rightarrow$~4$_{g}^{+}$           & 4.5 (5)  & 1.93 & 0.69 & +1.1 (3)                  & 9 (2) \\
996  & 4$_{\gamma}^{+}$~$\rightarrow$~4$_{g}^{+}$       & <1.4     & 1.10 & 0.45 & -                         & -     \\
570  & 4$_{\gamma'}^{+}$~$\rightarrow$~4$_{\gamma}^{+}$ & 7.8 (16) & 4.58 & 1.37 & -                         & >4.7 \\
1162 & 4$_{\gamma'}^{+}$~$\rightarrow$~4$_{g}^{+}$      & $\le$0.6 & 0.74 & 0.33 & -1.62$^{+0.76}_{-\infty}$ & <0.8  \\
664  & 6$_{\gamma}^{+}$~$\rightarrow$~6$_{g}^{+}$       & 1.3 (1)  & 3.09 & 1.00 & -1.01$^{+0.14}_{-0.18}$   & 0     \\
1002 & 6$_{g'}^{+}$~$\rightarrow$~6$_{g}^{+}$           & 1.1 (4)  & 1.08 & 0.44 & +1.08$^{+0.29}_{-0.25}$   & 1.5 (1.7) \\
\bottomrule
\bottomrule
\end{tabular}
\begin{tablenotes}
\item[a] From comparison with the $0_{g'} \xrightarrow{330} 2_{g}$ transition (see text).
\end{tablenotes}
\end{threeparttable}
}
\end{table}

%---------------------------------------------------------------------------------------
\section{Conclusions and future work}
\label{sec:conclusions}

Solenogam is an array of HPGe and Si(Li) detectors located at the focus of the SOLITAIRE recoil separator.  It was designed to study the decay of long-lived nuclear states, with a particular focus on extracting conversion coefficients.  Through a series of commissioning experiments, the system has demonstrated its ability to isolate long-lived states in a low-background environment and extract conversion coefficients using singles and coincidence spectra of both $\gamma$ rays and conversion electrons.  Attempts were made to study E0 transitions and extract both the E2/M1 and E0/E2 mixing ratios from conversion coefficients and/or angular correlations.  This was unsuccessful due to the difficulty of the e$^{-}$ measurements (electron statistics and energy resolution) and the limited sensitivity for $\gamma$-ray angular correlations.  In order to address these issues and improve the array in general, a number of upgrades are being discussed.  These include:

\begin{itemize}
\item Replacing the Si(Li) array with a larger number of smaller crystals.  The reduced surface area of the crystals leads to better energy resolution due to the decreased capacitance of the detectors~\cite{Knoll2000};
\item Changing the angular placement of the HPGe detectors to increase the sensitivity to angular correlation effects;
\item Adding a position-sensitive detector upstream of the Si(Li)~array to image the evaporation residues.  This will decrease the experimental setup time required to find the optimal field strength that focuses the evaporation residues, while also allowing  monitoring of the beam spot during experiments; and
\item Installing a moving tape system to decrease the system background by enabling the removal of long-lived contaminants from within the detector array.
\end{itemize}

The data acquisition system at the ANU-HIAF has also been recently upgraded to a digital system using XIA Pixie-16 digitisers~\cite{XIA2018} that can handle much higher data rates than the analogue system.  This allows the collection of greater statistics and provides more flexibility in offline analysis.

A more recent focus of the SOLENOGAM system has been the study of nuclear isomers in other regions of the nuclear chart.  Measurements of the high-spin isomer in $^{145}$Sm have been performed to better understand the systematic appearance of an isomeric, 49/2$^{+}$ state in the N=83 isotones~\cite{Odahara1997}.  Analysis of these data is ongoing, however the measurement of $\gamma$-$\gamma$ and $\gamma$-e$^{-}$ coincidences has already proven crucial to understanding the level scheme.  Further experiments on other isomers are also planned.

\section*{Acknowledgements}

The authors are grateful to the academic and technical staff of the Department of Nuclear Physics (Australian National University) and the Heavy Ion Accelerator Facility for their continued support.  This research was supported by the Australian Research Council through grant numbers FT100100991, DP120101417, DP14102986, and DP140103317.  M.S.M.G., A.A., B.J.C., J.T.H.D., T.J.G., B.Q.L., and T.P. acknowledge the support of the Australian Government Research Training Program.  Support for the ANU Heavy Ion Accelerator Facility operations through the Australian National Collaborative Research Infrastructure Strategy (NCRIS) program is acknowledged.

%\section*{References}
\bibliographystyle{elsarticle-num}
\bibliography{mybibfile}

\end{document}